\documentclass[apj]{emulateapj}
\usepackage{psfig}
 
\shortauthors{Mirabal et al.}
\shorttitle{X-ray Afterglow of GRB 970815}

\received{2004 August 10}
\accepted{2004 October 25}
\slugcomment{}

\begin{document}

\def\source{3EG~J1621+8203}
\def\grb{GRB~970815}
\def\axj{AX/RX J1606.8+8130}
\def\ro{{\it ROSAT\/}}
\def\asca{{\it ASCA\/}}
\def\beppo{{\it BeppoSAX\/}}
\def\rxte{{\it RXTE\/}}
\def\etal{{\it et al.}}
\def\ie{{i.e.}}
\def\eg{{e.g.,}}
\def\cha{{\it Chandra\/}}
\def\lsim{\mathrel{\lower .85ex\hbox{\rlap{$\sim$}\raise
.95ex\hbox{$<$} }}}
\def\gsim{\mathrel{\lower .80ex\hbox{\rlap{$\sim$}\raise
.90ex\hbox{$>$} }}}
\newbox\grsign \setbox\grsign=\hbox{$>$}
\newdimen\grdimen \grdimen=\ht\grsign
\newbox\laxbox \newbox\gaxbox
\setbox\gaxbox=\hbox{\raise.5ex\hbox{$>$}\llap
     {\lower.5ex\hbox{$\sim$}}}\ht1=\grdimen\dp1=0pt
\setbox\laxbox=\hbox{\raise.5ex\hbox{$<$}\llap
     {\lower.5ex\hbox{$\sim$}}}\ht2=\grdimen\dp2=0pt
\def\gax{\mathrel{\copy\gaxbox}}
\def\lax{\mathrel{\copy\laxbox}}
\def\pz{\phantom{0}}
\def\lsim{\mathrel{\lower .85ex\hbox{\rlap{$\sim$}\raise
.95ex\hbox{$<$} }}}
\def\gsim{\mathrel{\lower .80ex\hbox{\rlap{$\sim$}\raise
.90ex\hbox{$>$} }}}

\title{The X-ray Afterglow of Dark GRB 970815: A Common Origin for 
GRBs and XRFs?}

\author{N. Mirabal\altaffilmark{1,2}, J. P. Halpern\altaffilmark{1,2}, 
E. V. Gotthelf\altaffilmark{2}, and R. Mukherjee\altaffilmark{3}}
\altaffiltext{1}{Department of Astronomy, Columbia University,
  550 West 120th Street, New York, NY~10027}
\altaffiltext{2}{Columbia Astrophysics Laboratory, Columbia University,
  550 West 120th Street, New York, NY~10027}
\altaffiltext{3}{Department of Physics \& Astronomy, Barnard 
College, New York, NY~10027}

\begin{abstract}
\grb\ was a well-localized 
gamma-ray burst (GRB) detected by the All-Sky
Monitor (ASM) on the Rossi X-Ray Timing Explorer (\rxte) for which no 
afterglow was identified despite follow-up \asca\ and \ro\ pointings 
and optical imaging to limiting magnitude $R > 23$. While 
an X-ray source, \axj, 
was detected just outside the ASM error box, it was never
associated with the GRB because it was not clearly fading and because no 
optical afterglow was ever found.  We recently obtained an upper limit
for this source with \cha\ that is at least factor of 100 fainter than
the \asca\ detection.  We also made 
deep optical observations of the \axj\ position, which 
is blank to limits $V > 25.2$ and $I > 24.0$.
In view of these extreme limits we conclude 
that \axj\ was indeed the afterglow of
\grb, which corresponds to an optically ``dark'' GRB.
\axj\ can therefore be ruled out as the counterpart of the
persistent EGRET source \source.
The early light curves from BATSE and the \rxte\ ASM show
spectral softening between multiple peaks of prompt emission. 
We propose that \grb\ might be a case in which
the properties of an X-ray flash (XRF) and a ``normal'' 
GRB coincide in a single event. 
\end{abstract}

\keywords{
gamma rays: bursts --- gamma rays: observations --- X-rays: individual
(\grb)}

\section{Introduction}\label{sec:intro} 

One of the intriguing results from five years of
GRB localizations by \beppo\ is that roughly 60$\%$ of 
well-localized GRBs lack an optical transient despite intensive 
searches \citep[e.g.,][]{rei01,djo01}.
Some of these ``dark'' GRBs could simply be due to a 
failure to image deeply or quickly enough \citep{fox03,li03,lamb}.
However, in certain
cases the optical afterglow may have been missed either because
it is obscured by dust in the host galaxy,
or because it is located at high redshift ($z\gax5$). 

In the
first few months of the ``afterglow era'', which began with 
the localization of the X-ray afterglow of GRB 970228 \citep{cos97}, 
the Burst and Transient Source Experiment (BATSE) 
detected a GRB that falls in the category of ``dark''. The
bright event detected on UT 1997 August 15.50491 and 
labeled \grb\ had a total $\gamma$-ray fluence
$\approx 5.8 \times 10^{-5}$ erg cm$^{-2}$,
placing it in the top 15$\%$ of the BATSE fluence distribution. 
Nearly simultaneous detection by the
\rxte\ ASM refined the position of \grb\ to
a small error box \citep{smi97,smi99}. The localization by the
\rxte\ ASM was followed several days later by \asca\ \citep{mur97} 
and \ro\ \citep{gre97} pointings.  While a bright X-ray source \axj\
was detected just outside the ASM error box, it was never
associated with the GRB because it was not clearly fading 
and because prompt optical observations failed to reveal an 
optical transient to limiting magnitude $R > 23$ \citep{har97}.

In a subsequent review of the evidence we hypothesized nevertheless
that \axj\ was the afterglow of \grb, and proposed that this could be tested
\citep{mir03a}.
In this paper, we present new \cha\ and optical observations of
this source, which, together with an analysis of the \asca\ and \ro\
data, indicate that \grb\ was one of the earliest and most luminous
``dark'' bursts in the afterglow era [see \citet{dep03} for
a complete list].  In addition, we discuss 
the unusual softening over the burst's multiple peaks, which suggests that the
intrinsic properties GRB 970815 varied over the duration of the event. 
Finally, we mention
the implications for the counterpart of the steady unidentified EGRET 
source \source\ \citep{muk02}, 
whose error ellipse includes the position of \grb.

\begin{figure*}
\begin{center}
\epsscale{1.0}
\plottwo{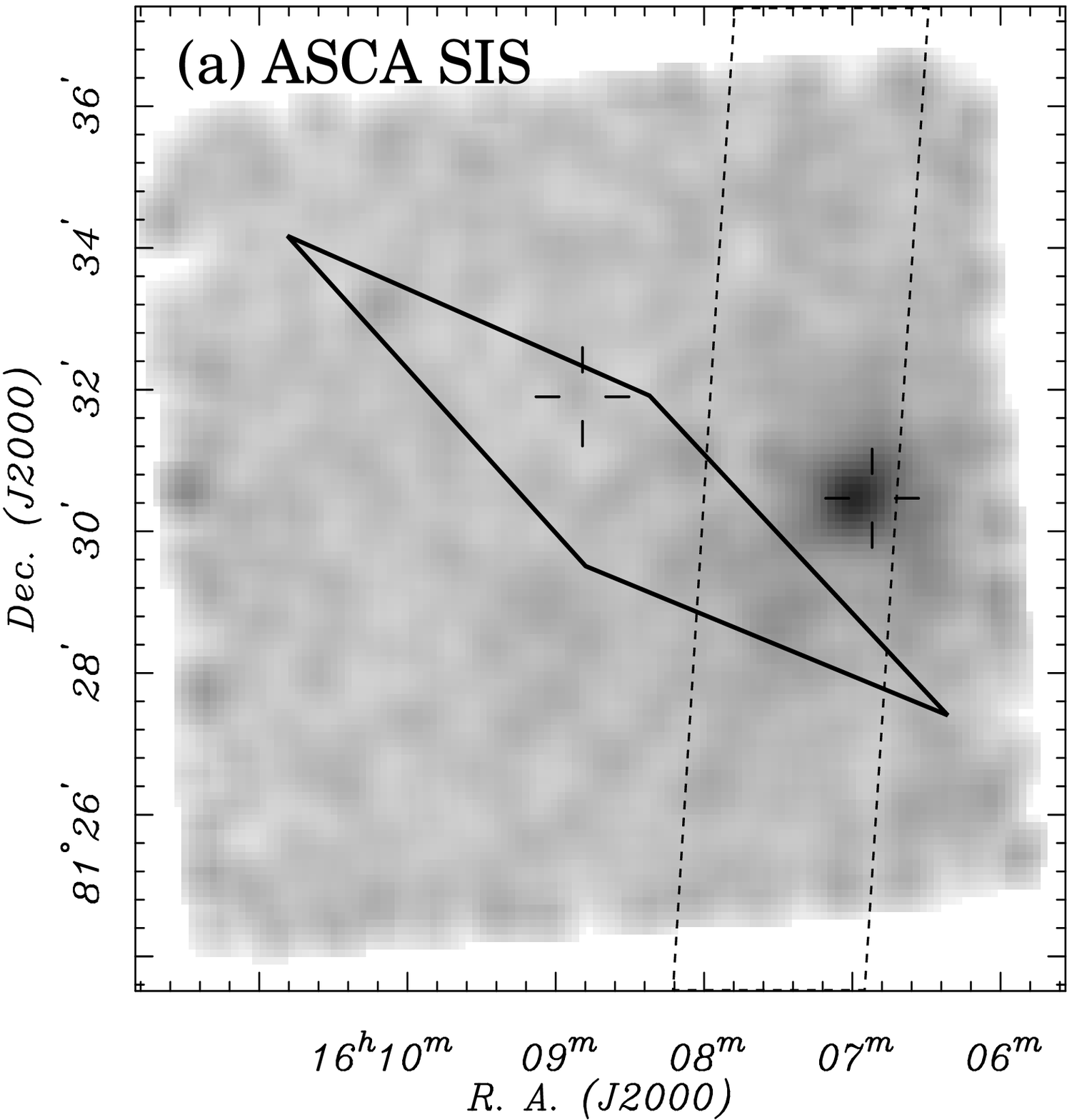}{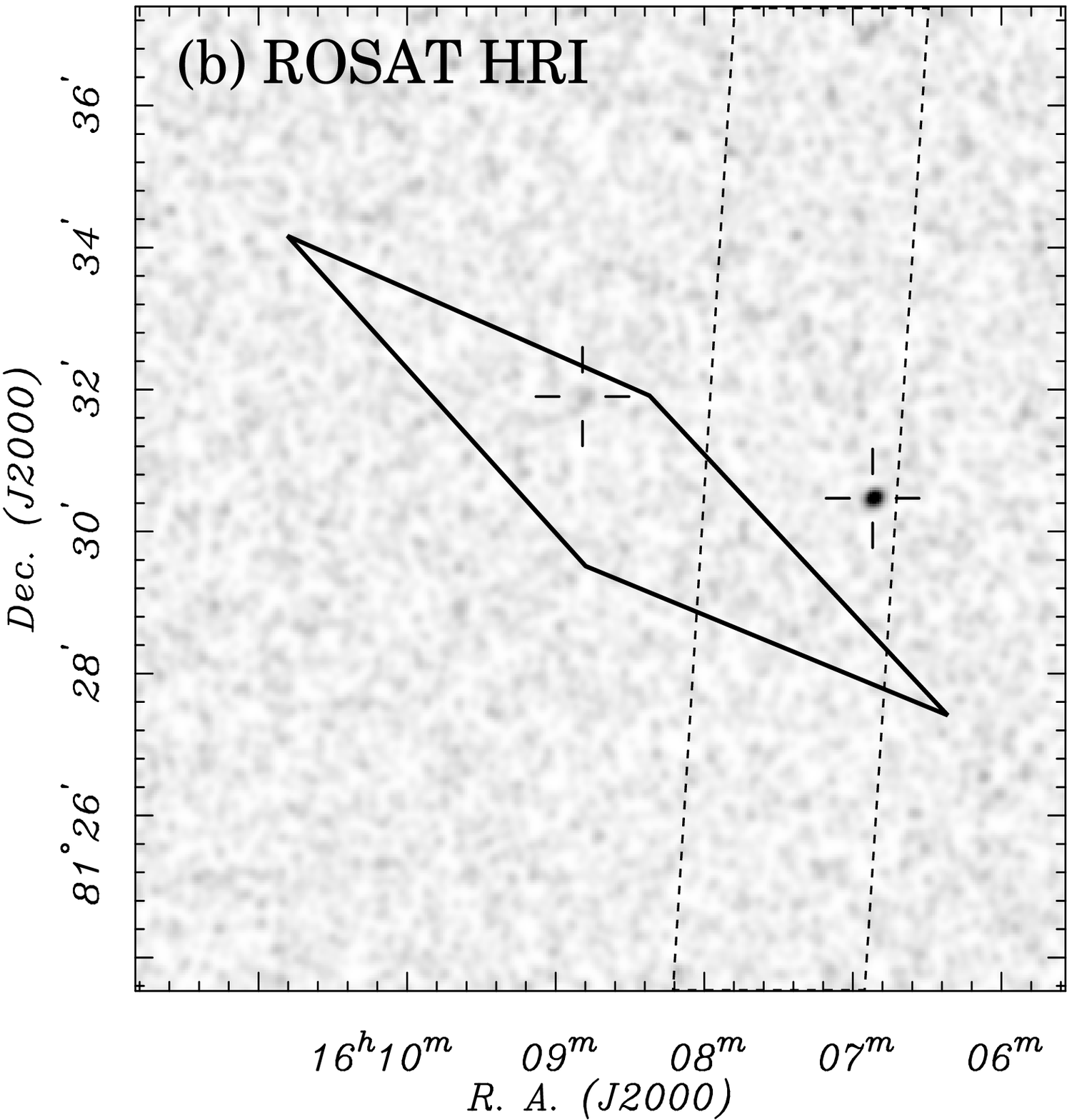}
\caption{
({\it a}) \asca\ SIS CCD image of the field of GRB 970815 at
3.2--4.4 days after the burst, with the \rxte\ ASM error box
({\it solid line}) and BATSE/{\it Ulysses} annulus 
({\it dashed lines}) from \citet{smi02}
superposed.  ({\it b}) \ro\ HRI image at
5.5--7.2 days after the burst.  Locations of \ro\ HRI point 
sources are indicated by crosses.  The marginal \ro\ source
RX~J1608.8+8131 \citep{gre97} is probably not real (see text).
\label{box}
}
\vskip -0.2in
\end{center}
\end{figure*}

\section{X-ray Observations}

\subsection{Prompt Localization and Follow-Up}

\grb\ was localized by the \rxte\ ASM on UT 1997
Aug. 15.50623 \citep{smi97}.
Simultaneous detection with two of the ASM scanning cameras
refined the position of \grb\ to
the small error box shown in Figure~\ref{box} \citep{smi99}.
The superposed annulus based 
on the BATSE and {\it Ulysses} triangulation 
confirmed the ASM position \citep{smi99}.
The prompt (1.5--12 keV) X-ray light curve had a multiple-peak 
structure lasting $\approx 130$~s, and
reaching a maximum intensity of $\approx 2$~Crab \citep{smi02}.

Following the prompt localization by \rxte, two   
X-ray observations were made that covered the entire
\rxte\ error box, 
one by \asca\ and one by the \ro\ High Resolution Imager (HRI). 
The \asca\ observation took place on
UT 1997 August 18.71--19.88, 
3.2--4.4 days after the burst \citep{mur97}, for a total usable
exposure time of 54.8 ks in both Gas Imaging Spectrometer (GIS) 
and Solid-state Imaging Spectrometer (SIS) detectors. 
Analysis of the data revealed no source brighter than
1 $\times 10^{-13}$~ergs~cm$^{-2}$~s$^{-1}$ within the \rxte\ error box.  
There was, however, a 
source  AX~J1606.8+8130 just outside the \rxte\
error box with an average flux 
$F_X$(2--10~keV) $= 4.2 \times 10^{-13}$~ergs~cm$^{-2}$~s$^{-1}$. 
Figure~\ref{box} shows the combined \asca\ SIS image and the location of
AX~J1606.8+8130 with respect to the burst error box.

The second X-ray observation of the \rxte\ error box
was obtained during UT 1997 August 20.99--22.73 with the \ro\ 
HRI, 5.5--7.2 days after the burst, with a total exposure time
of 17.1~ks \citep{gre97}. This observation (Fig.~\ref{box}) 
detected a source at (J2000.0) $16^{\rm h}06^{\rm m}52.\!^{\rm s}0, 
+81^{\circ}30^{\prime}28^{\prime\prime}$, consistent with
but more precise than the position 
of the \asca~source (hereafter referred to as \axj).  
The count rate $(3.4 \pm 0.5) \times 10^{-3}$~s$^{-1}$ 
extracted from a $15^{\prime\prime}$ radius centered
on RX~J1606.8+8130
corresponds to an extrapolated flux in the 2--10 keV band of
$2.1 \times 10^{-13}$~ergs~cm$^{-2}$~s$^{-1}$, or
$\approx 1/2$ the \asca\ value.  This extrapolation 
assumes the power-law spectral parameters derived in the next section
from the \asca\ source.
In addition, \citet{gre97} noted a fainter \ro\ source RX~J1608.8+8131 
with a flux of $\sim 5 \times 
10^{-14}$~ergs~cm$^{-2}$~s$^{-1}$ in the 0.1--2.4~keV band. 
This clouded the interpretation because,
although RX~J1608.8+8131 lies inside the \rxte\ error box,
its existence is of marginal statistical significance.
This possible source does not warrant further
comment, as it was not detected in the earlier \asca\ observation.
We concentrate our attention on the brighter source \axj\ which,  
although it lies just outside the \rxte\ error box,
is within the BATSE/{\it Ulysses} annulus.

\subsection{ASCA Spectral Parameters}

\begin{figure}
\begin{center}
\epsscale{0.80}
\psfig{file=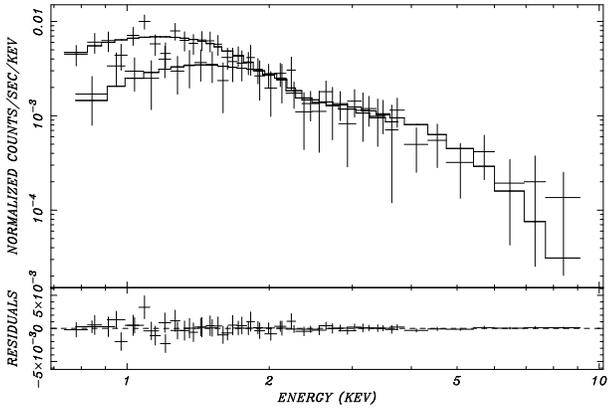,height=2.1in,angle=-90.}
\caption{
\asca\ GIS ({\it lower line}) and 
SIS ({\it upper line}) spectra of the source \axj.
{\it Top}: Data ({\it crosses})
and best-fit simultaneous absorbed power-law model
({\it solid line}), which has photon
index $\Gamma = 1.64 \pm 0.35$. 
{\it Bottom}: Difference between data and model;
units are the same as in the top panel.
\label{spec}
}
\vskip -0.4in
\end{center}
\end{figure}

The \asca\ GIS and SIS spectra 
of \axj\ are shown in Figure~\ref{spec}.
We fitted the spectra individually as well as 
jointly with common model parameters by
treating the normalization constant as a free parameter.
A simple absorbed power-law model provides a good description 
of the spectrum with photon index $\Gamma = 1.64 \pm 0.35$ and
$N_{\rm H} < 1.3 \times 10^{21}$~cm$^{-2}$ (the error bars 
corresponds to 90\% confidence for two interesting parameters). 
The fitted spectral index is insensitive to Galactic absorbing column
density whether $N_{\rm H}$ is treated as a free parameter or
held fixed at the maximum Galactic value in this direction,
$N_{\rm H, Gal} = 4.6 \times 10^{20}$~cm$^{-2}$. 

Since discrete X-ray
emission features have been reported in a few GRB afterglow
spectra \citep[see][]{pir00}, we looked for 
discrete emission
features, absorption edges and narrow radiative recombination
continua in the X-ray spectrum following the procedure 
described in \citet*{mir03b}. 
Unfortunately, the absence of
a redshift determination weakens the search. Thus, 
we proceeded to determine upper limits on equivalent width 
by holding
the power-law model parameters fixed and assuming a 
Gaussian line profile of fixed velocity width. The derived 
upper limit (90\% confidence level) 
corresponds to EW $< 0.2$~keV at 1.5~keV for a line of FWHM 
comparable to GRB 991216 \citep{pir00}.
This is less than than the reported EW measurement in GRB 991216,
so long as the redshift of \grb\ does not exceed $z \approx 1.3$.

\subsection{Chandra Observation}

The entire error box of \grb\ was observered on 2004 June~17 with the
Advanced CCD Imaging Spectrometer \citep[ACIS;][]{bur97} onboard the
\cha\ X-ray Observatory \citep*{wei96,wei02}.  The source \axj\ was
positioned at the default location on the back-illuminated S3 CCD of
the ACIS-S array.  The standard TIMED readout with a frame time
of 3.2~s was used, and the data were
collected in VFAINT mode.  A total of 10130~s of on-time was
accumulated, while the effective exposure live-time was 9998~s.
We verified that the \cha\ astrometry is accurate to $0.\!^{\prime\prime}3$
or better in each coordinate by identifying four serendipitous sources
on our optical images.
Within the $10^{\prime\prime}$ radius error circle of \axj,
there is no \cha\ point source with more than one photon in the 
0.2--10~keV band.  Adopting a 96\% confidence upper limit of five photons,
we convert to a flux upper limit in the 2--10~keV band using the
\asca\ spectral index $\Gamma = 1.64$ and
$N_{\rm H} < 1.3 \times 10^{21}$~cm$^{-2}$.
The Web-based simulator
PIMMS\footnote{Available at http://asc.harvard.edu/toolkit/pimms.jsp.}
allows us to make this conversion while
accounting for the time-dependent degradation of the ACIS throughput
in the AO5 observing period in which the observation was conducted.
The result is $ F_X(2-10\,{\rm keV}) < 3.7 \times 10^{-15}$ ergs~cm$^{-2}$~s$^{-1}$,
or less than 1\% of the \asca\ measured flux in the
same band.  Such a dramatic disappearance is strong evidence that \axj\
was the afterglow of \grb.  In combination with the lack of an optical
counterpart such as a variable star or galactic nucleus (see below),
this identification is compelling.

We also note that nothing was detected by \cha\ at the location of the 
marginal \ro\ source RX~J1608.8+8131 \citep{gre97} to a similar flux limit.  
In the absence of any other evidence for the existence of this
source, we conclude that it was never real.

\begin{figure}
\begin{center}
\epsscale{1.05}
\plotone{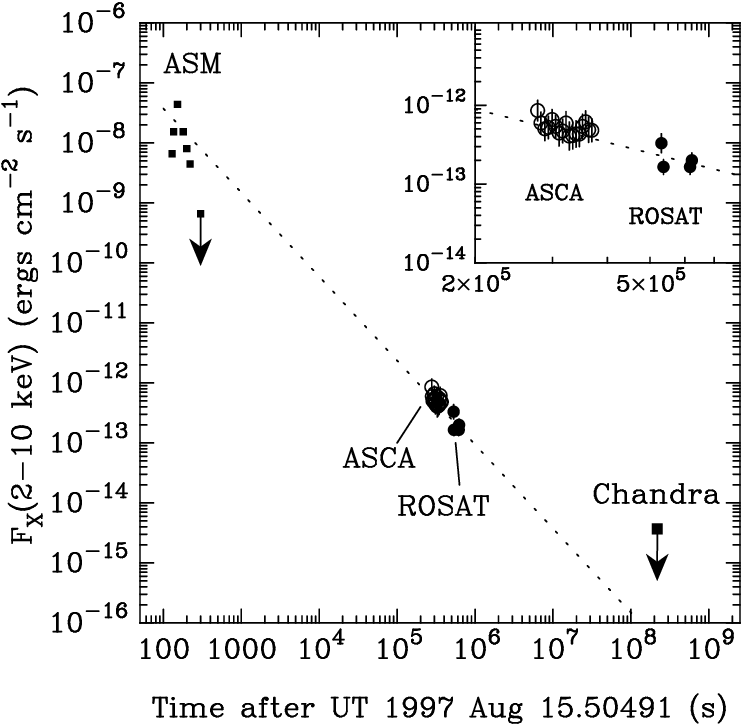}
\caption{
The X-ray light curve of the \grb\ afterglow.
\rxte\ ASM fluxes were derived by converting the reported 1.5--12 keV 
power-law spectrum \citep{smi02} to the 2--10 keV energy band.
The arrows indicate ASM and \cha\ upper limits.
\ro\ HRI fluxes were derived by assuming that the
source has the same power-law spectrum as its \asca\ counterpart.
The dotted line shows a power-law
decay $F_X \propto t^{-1.4}$, although the
variation in the \asca\ points are also consistent
with no overall decay. 
{\it Inset\/}: Expanded view of the \asca\ and \ro\ light curves.
\label{lc}
}
\vskip -0.2in
\end{center}
\end{figure}

\subsection{Combined X-ray Light Curve}

Figure~\ref{lc} shows the combined X-ray light curve of \grb.
Comparison of the various energy channels
of the ASM and BATSE indicates that the third and final
peak in the ASM (1.5--12 keV) prompt emission, the one that
began $\approx 130$ after the BATSE trigger, has the softest
spectrum with a peak energy in $\nu F_{\nu}$ of 
$E_{\rm peak} \leq 25$ keV and a photon index
$\Gamma = 1.8 \pm 0.1$ \citep{smi02}.  The latter authors
suggested that this third peak is
the beginning of the afterglow phase as a relativistic 
shock decelerates. The flux during the third peak,
converted here from the reported ASM flux to
the 2--10 keV energy band, reached a maximum
$F_X(2-10\,{\rm keV}) = 4.4 \times 10^{-8}$ ergs cm$^{-2}$ s$^{-1}$ 
( $\approx 2$~Crab) at 
$t = 152$~s after the BATSE trigger \citep{smi02}. It then dimmed 
drastically during the next 148~s to $F_X(2-10\,{\rm keV}) \leq 
6.6 \times 10^{-10}$ ergs cm$^{-2}$ s$^{-1}$ \citep{smi02}.
Fitting the ASM points to a power law requires a
decay as steep as $F_X \propto t^{-6.2}$ with the origin
of time set at the BATSE trigger.  We show this early decay phase
in Figure~\ref{lc}.

The \asca\ light curve in Figure~\ref{lc} consists of
the sum of the counts from all four of its detectors. The \ro\ points 
correspond to an extrapolated flux in the 2--10~keV band 
assuming the power-law spectral parameters derived from \asca.
The individual \asca\ and \ro\ components of the light curve
show no obvious evidence for variability.  However, if
the flux remained constant between the \asca\ and
\ro\ observation, then we should
expect to find a total of $\approx 114$ source 
photons in the 0.1$-$2.0~keV \ro\ energy band, whereas only 63 net 
photons are detected in the HRI 
observation. The Poisson probability of obtaining 63 or fewer events
when 114 are expected is $1.3 \times 10^{-7}$.
Instead, we find that the flux
of \axj\ is more consistent with a 
$F_X \propto t^{-1.4}$ decay between the \asca\ and \ro\ 
observations, easily within the range of well-studied 
GRB X-ray afterglows.   If we extrapolate the 
2--10 keV X-ray flux from 500~s to $10^6$~s after the burst 
using $\alpha = -1.4$, 
we get a fluence of  $4.4 \times 10^{-6}$ ergs~cm$^{-2}$ or 
$\approx 8\%$ of the burst fluence, which is in agreement
with the properties of other GRBs \citep{fro00}.  

\begin{figure}
\begin{center}
\epsscale{0.9}
\plotone{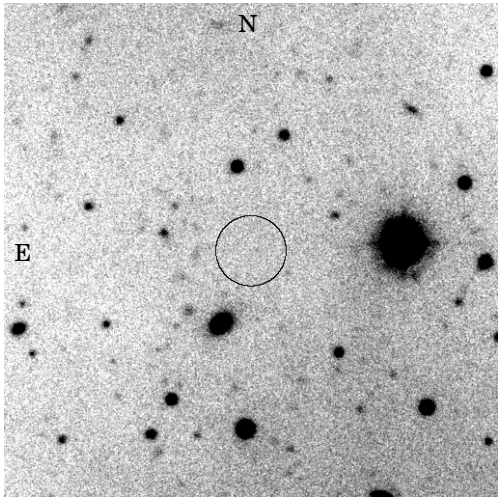}
\caption{
A $V$-band image taken on 2004 July~12 with the
MDM Observatory 2.4~m telescope at the location of the unidentified
X-ray source \axj, whose \ro\ HRI position is (J2000.0)
$16^{\rm h}06^{\rm m}52.\!^{\rm s}0, +81^{\circ}30^{\prime}28^{\prime\prime}$.
The seeing was $1.\!^{\prime\prime}6$.
The field is $2^{\prime}\!.3$ across,
and the adopted \ro\ HRI error circle is
a conservative $10^{\prime\prime}$ in radius.
The 3$\sigma$ upper limit is $V > 25.2$.
\label{opt}
}
\vskip -0.2in
\end{center}
\end{figure}

\section{Optical and Radio Observations of AX/RX~J1606.8+8130}

Following the rapid dissemination of the \rxte\ position for \grb,
a number of groups obtained
optical images of its error box including the position of \axj.
No significant variable source was found in or near the \rxte\ error box
to limits $V > 21.5$ \citep{gro97}, $R > 21$ \citep*{sta97}, and 
$R > 23$ \citep{har97} starting 14--17 hours after the burst.
Much later, while conducting a 
search for the $\gamma$-ray source \source\ \citep{muk02},
we obtained deep optical images in several colors of the X-ray position
of \axj\ with the MDM Observatory 1.3~m and 2.4~m telescopes
over the period 2001 June -- 2004 July. 
Figure~\ref{opt} shows a $V$-band image
obtained on 2004 July~12.
The adopted $10^{\prime\prime}$ radius \ro\ error
circle is conservative, since the \ro\ aspect
is confirmed by the detection of the bright star BD+82~477 in the same
image \citep{muk02}.  The error circle is blank to a 3$\sigma$ limit
of $V > 25.2$, which corresponds to $F_X/F_V > 800$ for the
\asca\ source under the definition of \cite{mac82}.
In other filters, \axj\ shows no evidence of a host galaxy or
any other optical counterpart to
limits of $B > 21.5$, $R > 22.0$, and $I > 24.0$. 
Such extreme $F_X/F_V$ ratios are seen only among isolated neutron 
stars or low-mass X-ray binaries.  The former is ruled out here by 
the extreme X-ray variability, and the latter by the absence of an 
optical counterpart.  Thus, we are convinced that the X-ray afterglow
of \grb\ was detected.

Several non-detections were obtained with the VLA between 1 and 103 days
after the burst at frequencies of 4.89 and 8.44~GHz \citep{fra03}.
The rms noise in these observations ranged from 98 to 16 $\mu$Jy.

\section{Discussion}

\subsection{\grb\ as a Dark Burst}

Although the \asca/\ro\ light curve supports a possible decay for
\axj\ \citep{gre97}, the follow-up efforts for \grb\
were abandoned prematurely, we judge in hindsight, mainly 
because of the small positional 
inconsistency of \axj\ with the \rxte\ error box, and 
the absence of an optical afterglow. 
Little was known about ``dark'' GRBs at the time
to motivate further observations.
In fact, the ``dark'' GRB hypothesis is justified when
one extrapolates the X-ray decay and spectral index backward to predict the  
optical magnitude at the time of the reported optical observations.
It is important to note that there are now 
many examples of non-monotonic decays in GRB afterglows; therefore,
the observed behavior of \grb\ may not be representative
of its long-term decay. 
However, the following analysis is reasonable as long as the deviations are not 
extreme. Starting with  
the observed X-ray flux density $f_{X}$,  we can extrapolate a broad-band
spectrum of the form $f_{R}
= f_{X}(\nu_{R}/\nu_X)^{-\beta}$ where $f_{R}$ is the $R$-band 
optical flux density at a frequency $\nu_{R}$ and $\beta$ is the X-ray
spectral index. 
From the \asca\ spectra we have $f_{X} \approx 0.10\,\mu$Jy 
($\nu_{X}=4.84 \times 10^{17}$~Hz) at a time 
$t \approx 3.74$ days after the burst, and $\beta \approx
0.64$.  The optical flux density evolution would then correspond to
$f_{R}(t_d) \approx 55\,t_{d}^{-1.4}\,\mu$Jy
where $t_d$ is days
elapsed since the BATSE trigger. This translates into 
$R \approx 19.0$ on UT 1997 August~16.31. 
Therefore, the predicted magnitude is brighter than the
$R > 21$ \citep{sta97}, or  $R > 23$ \citep{har97}
upper limits reported at that time.
The difference would require an observer-frame extinction
$A_R \gax$ 4 mag. 

In order to convert the observer-frame extinction to the rest frame of
the host galaxy, we make a simple assumption that its redshift
falls near the average GRB redshift, $<\!z\!> \approx 1.4$.
This is a conservative assumption for the sake of our argument, since 
the required rest-frame extinction increases if $z < 1.4$. 
At $z \approx 1.4$, the effective $R$-band wavelength is
$\approx 2740$ \AA. Assuming an extinction curve with 
a fixed form \citep{car89},
this translates into a visual extinction $A_{V,{\rm rest}} \gax$ 2 mag. 
A rest-frame extinction $A_{V,{\rm rest}} \gax 2$ for $z \lax 1.4$ 
implies significant dust extinction at the host galaxy, possibly
characteristic of molecular clouds at the birth site of the GRB progenitor 
\citep{djo01}, and supports a ``dark'' GRB description.

Based on the plausible values of observed column density
($N_{\rm H} < 1.3 \times 10^{21}$~cm$^{-2}$), 
we cannot formally rule out 
large extinction at the host galaxy from the X-ray spectra alone. 
In fact, this maximum allowed column density (90$\%$ confidence 
level) would translate to 
$N_{\rm H,rest} \approx 10^{22}$ cm$^{-2}$ at $z \approx 1.4$ \citep{mor83}.
The derived $N_{\rm H,rest}$ is well 
within the characteristic column density for giant molecular clouds 
found in our Galaxy \citep{sol87}.
The values obtained for $z \lax 1.4$ are also
in rough agreement with the relation between $A_{V}$ and $N_{\rm H}$ for the
Milky Way \citep{pre95}. It is possible that  
effects such as dust sublimation \citep{wax00}
and  grain charging \citep*{fru01} can play a 
significant role in GRB environments. These
dust destruction mechanisms could be effective as far
as $\sim$ 100 pc from the burst site, which might lead 
to gray dust \citep[e.g.,][]{mir02} and lower extinction 
(Galama \& Wijers 2001).  Alternatively, the absence of an
optical afterglow could be attributed to a high redshift
$(z \gax 5)$ for which the Lyman break
moves into the $R$ passband. However, if interpreted as a jet at $z \gax 5$, 
GRB 970815 would require a very small 
opening angle $\theta_{\rm j} \leq 0.\!^{\circ}7$, once corrected for
a standard energy reservoir \citep{bloom}. Such a small angle might be
difficult to achieve in an expanding jet breaking through the circumburst 
medium. 

\subsection{Modeling the Afterglow and Reflecting on the Prompt Emission}

Of the synchrotron models involving a blast wave expanding relativistically
 in a   stellar-wind medium \citep{che99}, the 
  combination of electron power-law 
 distribution index $p = 2.2$, spectral index 
  $\beta$ = $(1 - p)/2 = -0.6$, and decay slope 
  $\alpha = (1 - 3p)/4 = -1.4$, corresponding to 
$\nu_{m} < \nu < \nu_{c}$, provides a remarkably good description for the 
  afterglow as measured by \asca\ and 
  \ro. Such a model, however, cannot account for the 
significantly steeper decay index ($\alpha = -6.2$) 
in the ASM light curve (Fig. ~\ref{lc}).  
One possibility is that
 the steepening in the decay follows the passage of the typical frequency
 $\nu_{m}$ through the X-ray band. However, this transition 
should steepen to 
 $\alpha = (2 - 3p)/4$ \citep{granot}, which yields a physically 
 unreasonable $p = 9$. Similar theoretical predictions 
 for the decay of reverse shock emission  
 impose an equally extreme $p \approx 8$ \citep{koba}. This led 
\citet{smi02} to suggest that the final peak might be due to refreshed shocks 
or density inhomogeneities. 
It is, however, difficult to reconcile a steep decay with 
energy or density variations \citep{nakar}. 
Thus, by a process of elimination, we find it
unlikely that the ASM data represent the beginning of the afterglow.

Instead, we propose that the third peak represents 
a continuation of the prompt GRB emission and the onset of 
a soft XRF. The latter are believed to arise from
a softer GRB mechanism that produces a peak energy
of order 1~keV $\leq E_{\rm peak} \leq$ 40~keV 
\citep{heise,kippen,sakamoto}.  Remarkably,  
the observed peaks in GRB 970815 drift by a large factor during the
duration of the burst, reaching a first maximum  with
$E_{\rm peak} \geq 110$ keV at $t \approx$ 1~s, another at $t \approx$ 98~s 
in the 
$60 \leq E_{\rm peak} \leq 110$ keV range, and a pronounced third
with $E_{\rm peak} \leq 25$ keV at $t \approx$ 152~s, in which 
an 8~s delay between the 
maximum in the C band ($E \approx 7$~keV) and the A band 
($E \approx 2.25$~keV) is observed \citep{smi02,bradt}. Interestingly, 
the third peak has a duration
($\approx$ 80~s) and power-law spectrum ($\Gamma = 1.8$)
comparable to the parameters of XRFs measured by
\beppo, BATSE and {\it HETE--2\/} \citep{heise,kippen,barraud,sakamoto}.
This might be an indication that 
the individual properties of an XRF and a ``normal'' 
GRB can coincide in a single event.
A possibly related phenomenon is the hard-to-soft spectral evolution
that has been seen in a number of \beppo\ and {\it HETE--2\/}
GRBs \citep[\eg][]{sakamoto}.
In addition, the precursors and tails of some GRBs seen by {\it Ginga}
had spectral properties similar to XRFs 
[see \citet{murakami} and references therein].
Since the prompt emission is a function of various physical parameters, 
it is unclear what provides the necessary softening over
multiple peaks. However, 
a variable Lorentz factor in a long-lived, ``tired'' central engine, 
or a decreasing magnetic field are attractive 
possibilities \citep{lloyd}.

\subsection{Implications for the Counterpart of 3EG~J1621+8203}

Our analysis of \axj\ also has implications 
for the completeness of the survey for a counterpart
of the unidentified EGRET source \source\ \citep{muk02}, 
whose error ellipse includes the position of \grb. 
Based on existing X-ray and radio data, the FR~I radio galaxy
NGC~6251 ranks as the most likely counterpart for \source\ 
\citep{muk02} now that \axj\ has been eliminated from consideration.
NGC~6251 is a notable object because of the possible link between BL~Lac 
objects and FR I radio galaxies. FR~I radio galaxies are
hypothesized to be the likely parent populations of BL~Lac objects, which are
believed to be FR~I radio galaxies with jets pointing near the line of sight
\citep{urr95}. In the Third EGRET catalog \citep{har99}
Cen~A (NGC~5128) is the only FR~I radio galaxy identified as a 
source at energies above 100~MeV \citep{sre99}.  
NGC~6251 could then be the
second FR~I radio galaxy to be detected in high-energy $\gamma$-rays. 

\section{Conclusions and Future Work}

In summary, our verification of the transient nature of \axj\ and the
lack of an optical counterpart for it compel the conclusion that
\grb\ was an optically ``dark'' GRB, quite possibly the
first one in the afterglow era. 
Its light curve can be fitted by a power-law decay of index $\alpha = -1.4$
between the \asca\ and \ro\ observations, 
with a spectrum of photon index $\Gamma = 1.64 \pm 0.35$.
Analysis of the \rxte\ ASM 
observation leads to the conclusion that at least some 
GRBs exhibit properties that are similar to XRFs after the cessation
of ``normal'' gamma-ray activity. Such a detection  
suggests that variations in the intrinsic properties of the burst 
might account partly for the observed distribution of XRFs and GRBs. 

This finding warrants a fresh examination of archival optical/IR data,
as well as follow-up optical, IR, and sub-mm observations
to search more deeply for a host galaxy and determine if it is 
obscured by dust or located at high redshift. 
Even if ambiguous within the \ro\ HRI error circle, identification
of the host may be supported by spectroscopic detection of
strong Ly$\alpha$ emission, a common signature of large
star formation rates in GRB host galaxies \citep{fynbo}.

\medskip

\acknowledgments 
We thank the referee for a number of valuable comments. 
We acknowledge helpful correspondence with Marc Kippen regarding BATSE.
This work was supported by SAO grant GO4-5057X, and by the National
Science Foundation under Grant. No. 0206051.

\end{document}